**Low-energy electron-irradiation effect on transport properties of graphene field effect transistor.**

F. Giubileo[1], A. Di Bartolomeo[1,2], N. Martucciello[1], F. Romeo[2,1], L. Iemmo[1,2], P. Romano[3] and M. Passacantando[4].

[1]*CNR-SPIN Salerno, via Giovanni Paolo II, 132, 84084, Fisciano, Italy*

[2]*Dipartimento di Fisica "E.R. Caianiello", Università di Salerno, via Giovanni Paolo II, 132, 84084, Fisciano, Italy*

[3]*Dipartimento di Scienze e Tecnologie, Università del Sannio, via Port'Arsa 11, Benevento, Italy*

[4]*Dipartimento di Scienze Fisiche e Chimiche, Università dell'Aquila, Via Vetoio, 67100, L'Aquila, Italy*

**Abstract**

We study the effects of low-energy electron beam irradiation up to 10 keV on graphene based field effect transistors. We fabricate metallic bilayer electrodes to contact mono- and bi-layer graphene flakes on $SiO_2$, obtaining specific contact resistivity $\rho_c \approx 19\ k\Omega\ \mu m^2$ and carrier mobility as high as 4000 $cm^2V^{-1}s^{-1}$. By using a highly doped *p*-Si/$SiO_2$ substrate as back gate, we analyze the transport properties of the device and the dependence on the pressure and on the electron bombardment. We demonstrate that low energy irradiation is detrimental on the transistor current capability, resulting in an increase of the contact resistance and a reduction of the carrier mobility even at electron doses as low as 30 $e^-/nm^2$. We also show that the irradiated devices recover by returning to their pristine state after few repeated electrical measurements.

**1. Introduction**

Graphene is a promising candidate for future nanoelectronics and has been attracting enormous attention by the scientific community since 2004, when graphene flakes were exfoliated from graphite for the first time in Manchester [1]. Post-silicon era seems to be close due to physical limits of Si-technology down-scaling and carbon-based electronics is considered as a possible option [2]. Carbon nanotubes (CNTs) have been largely studied in last two decades but two principal drawbacks are limiting their applicability: not-controllable chirality causing both metallic and semiconducting nanotubes in fabrication processes and difficulty to correctly place large number of nanotubes in integrated circuits. Graphene has reignited the idea of a carbon-based electronics offering unmatched properties as the linear dispersion relation with electrons behaving as massless Dirac fermions [3], the very high carrier mobility [4] and the superior current density capability [5]. Graphene in applications as gas sensors [6], photodetectors [7], solar cells [8], heterojunctions [9] and field-effect transistors [10] is already a reality.

From an experimental viewpoint, the use of scanning electron microscopy (SEM), transmission electron microscopy (TEM) as well as electron beam lithography (EBL) and focus ion beam (FIB) processing in ultra-high vacuum represents a necessary step for the fabrication and characterization of graphene based devices. Consequently, graphene devices during fabrication or under test are necessarily exposed to high vacuum and electron irradiation that may considerably affect their electronic properties.

Several experiments have shown that irradiation of energetic particles, such as electrons [11-15] and ions [16,17], can induce defects and damages in graphene and cause severe modifications of its properties. Raman spectroscopy has been largely used to study electron-beam induced structural modifications [18-20], or formation of nanocrystalline and amorphous carbon [17,21], as well as to correlate the reduction of 1/f noise in graphene devices with the increasing concentration of defects [22]. The shape and relative magnitude of a D peak as well as the shift of the G peak has been used to quantitatively evaluate the damage and the strain induced by very low energy e-beam [23]. Raman and Auger electron spectroscopy have proved that e-beam irradiation can selectively remove graphene layers and induce chemical reactions and structural transformations [19,20]. Interaction of e-beam with water adsorbates on the graphene surface has been also proposed for hydrogenation of graphene [24,25].

However, the Raman spectroscopy is unable to reveal all the effects of the e-beam irradiation, and electrical measurements are needed to check for possible modifications of transport properties. Despite that, electronic transport properties of irradiated graphene devices have not been deeply investigated as yet [26,27]. The negative shift of the Dirac point has been reported as effect of e-beam induced *n*-doping. The comparison with the case of suspended graphene, has evidenced also the importance of the substrate [26]: it has been demonstrated in particular that e-beam irradiation of graphene field effect transistors (GFETs) modifies the substrate band bending and results in localized *n*-doping of graphene which creates graphene *p-n* junctions working as photovoltaic device [28].

In this paper, we study the modification of electronic transport properties of GFETs upon exposure to electron beam irradiation for scanning electron microscopy imaging with acceleration energy up to 10 keV. An optimized fabrication process has been developed to obtain devices characterized by specific contact resistivity $\rho_c \approx 19 k\Omega\mu m^2$ and carrier mobility as high as 4000 cm$^2$V$^{-1}$s$^{-1}$ on Si/SiO$_2$ substrate. Electron irradiation affects the transistor current drive capability by reducing the carrier mobility and increasing the channel and contact resistance. We also show that for low energy electron irradiation the conditions of pristine devices are almost restored by successive gate voltage sweeps while measuring the channel conductance.

## 2. Experiment

Graphene flakes were obtained from highly oriented pyrolytic graphite by scotch tape method and were placed on standard *p*-Si SiO$_2$ (300nm thick) substrates. After optical identification, the mono- or bi-layer nature of the flakes were confirmed by Raman spectroscopy. Metal contacts to selected graphene flakes were realized by means of electron beam lithography (EBL) and magnetron sputtering techniques. Spin coating of approximately 400 nm PMMA-A7 (Poly-methyl methacrylate) at 4000 rpm was made on the sample, and it was successively exposed by Raith EBL system. Methyl isobutyl ketone and then isopropanol has been used as developer. The metal electrodes were fabricated by a three cathode RF Sputtering Magnetron for in-situ multilayer deposition working at 10$^{-7}$ mbar base pressure. The graphene flakes were contacted by Nb/Au metallic bilayer (15 nm Nb/25 nm Au) with niobium contacting the graphene and gold working as cap layer to prevent Nb oxidation and favor electrical connection with the probe tips. Metallic leads were sputtered at low power density (< 0.7 W cm$^{-2}$) and small deposition rates (0.3 nm/s for Nb and 1.2 nm/s for Au) to prevent graphene damages.

Electrical characterization was performed by means of a Janis Research ST-500 cryogenic probe station connected to a Keithley 4200 Semiconductor Characterization System (SCS) working in wide ranges of current (100 fA to 0.1 A) and voltage (10 µV to 200 V). To study the effect of e-beam irradiation on transistors, the SCS was connected to a scanning electron microscope equipped with Kleindeik nanomanipulators, which allowed in-situ electrical measurements with the sample inside the high-vacuum SEM chamber to prevent adsorbate contamination.

## 3. Results
### 3.1. Contact resistance

In order to characterize the contact resistance, we designed a device with standard geometry to apply the Transfer Length Method (TLM), the structure consisting of a series of spaced electrodes, up to 10 µm apart (figure 1a).

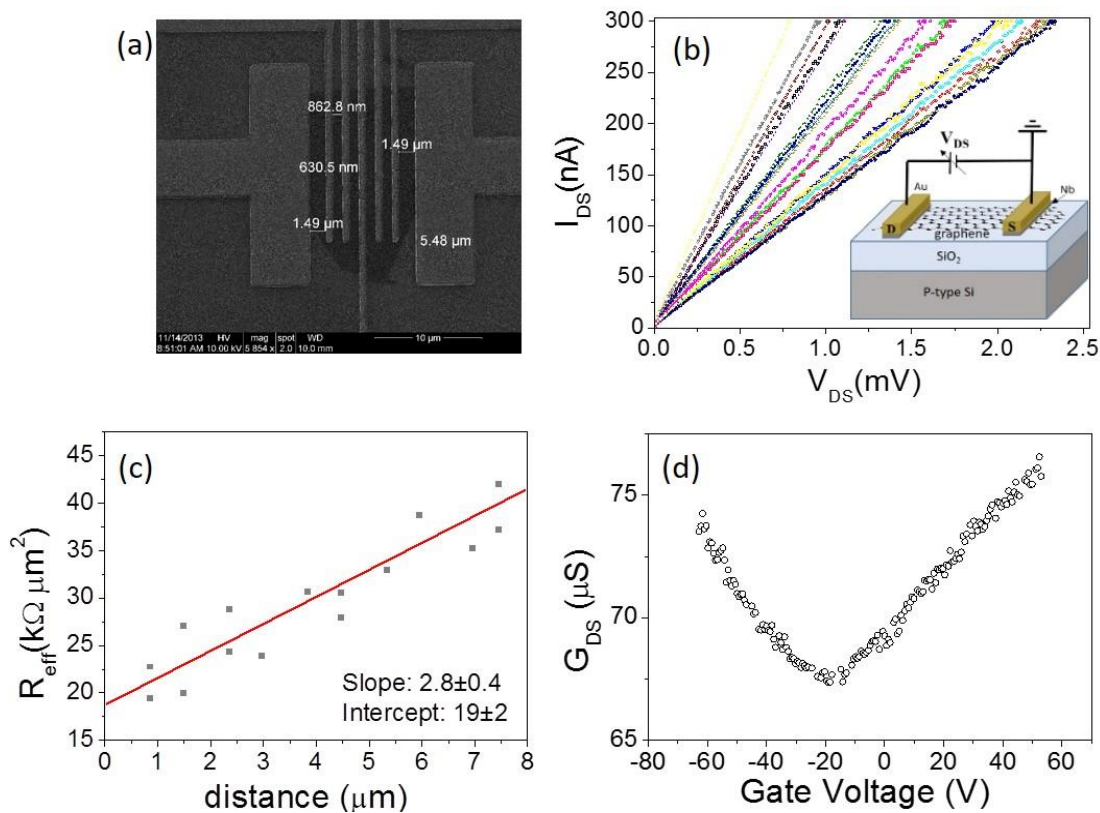

*Figure 1. a) A TLM device with Nb(15 nm)/Au(25 nm) contacts. b) Current-voltage characteristics measured between every two-contacts combination on the graphene flake; Inset: scheme of the device; c) TLM plot of $R_{eff}(L)$ at $V_{Gate} = 0V$; d) Transfer characteristic of one of the back-gated transistors of a) in the range -60 V < $V_{Gate}$ < +60 V.*

The two-probe current-voltage characteristic of the channel ($I_{DS}$ vs $V_{DS}$) has been measured for each possible combination of contacts: the drain current ($I_{DS}$) linearly increases with source-drain voltage ($V_{DS}$) which is a typical behavior at low bias (Figure 1b). According to the TLM, we can extract the specific contact resistivity $\rho_c$ by evaluating (for the general situation of irregular shaped flakes) the intercept of a plot of $R_{eff}$ vs $L$ [29], with $L$ the separation between the two electrodes, and

$$R_{eff} = R \left( \frac{1}{W_1 d_1} + \frac{1}{W_2 d_2} \right)^{-1}$$

where $W_i$ and $d_i$ (for $i = 1,2$) indicate width and length of each contact respectively. From the linear fitting of $R_{eff}$ vs $L$ (see Figure 1c), we find $\rho_c = 19 \pm 2\ k\Omega\ \mu m^2$, an intermediate value compared to previously reported values of $7\ k\Omega\ \mu m^2$ for Ni and $30\ k\Omega\ \mu m^2$ for Ti [30].

We also tested the current modulation of this device when used as field effect transistor with the Si substrate as the back-gate electrode. In figure 1d we report the transfer characteristic $G_{DS}$ vs $V_{Gate}$ in which the channel conductance $G_{DS}$ is measured as a function of the gate voltage $V_{Gate}$ between a couple of electrodes biased at $V_{DS}$ = 0.5 mV. The conductance clearly shows a minimum at $V_{Gate}$ = -15 V corresponding to the charge neutrality point (Dirac point). The negative value indicates that the graphene is *n*-doped. The device was measured as produced, without any electrical annealing (stress), that is suitable to induce desorption of surface contaminants as well as to improve the metal-graphene coupling, thus reducing the contact resistance [5]. In figure 2 we show the output characteristics ($I_{DS}$ vs $V_{DS}$ for several $V_{Gate}$ values in the range -60 V to +60 V) and the transfer characteristic (at fixed $V_{DS}$) measured before and after an electrical stress event that stabilize the device improving its performances. The black arrow in the figure identifies the voltage at which the device is suddenly modified, switching from a total resistance of about 250 $k\Omega$ to 150 $k\Omega$, for effect of current annealing. After such modification, the device has been routinely measured, showing extreme stability without further modification of the total resistance $R_{DS}$, which we report as a function of $V_{Gate}$ in the insets of figure 2. $R_{DS}$ is the series of the contact resistance and the channel resistance, $R_{DS} = R_c + R_{channel}$, where the channel resistance can be expressed as $R_{channel} = \frac{L/W}{\mu\, n(V_{bg})\, q}$ with $L$ and $W$ the length and width of the channel, respectively, $\mu$ is the carrier mobility, and $q$ is the unit charge [31]. The total carrier concentration can be written $(V_{bg}^*) = \sqrt{n_{ind}^2 + n_0^2}$, where $V_{bg}^*$ is the back gate voltage with respect the Dirac voltage ($V_{bg}^* = V_{bg} - V_{Dirac}$), $n_0$ is the intrinsic carrier concentration and $n_{ind}$ is the carrier concentration induced by the back gate. $n_{ind}$ can be expressed in terms of gate oxide capacitance as $n_{ind}(V_{bg}^*) = C_{ox} V_{bg}^*/q$. This model, adapted to the experimental data $R$ vs $V_{Gate}$, allows to extract the contact resistance and carrier mobility as fitting parameters.

Using the transfer characteristics measured before and after the electrical stress, we found that the contact resistance is improved (reduced from 200$k\Omega$ to 90 $k\Omega$), while the carrier mobility is increased from 3600 V$^2$cm$^{-1}$s$^{-1}$ to 3900 V$^2$cm$^{-1}$s$^{-1}$. The electrical stress increases the graphene–metal coupling and acts as cleaning of the channel. The mobility values are comparable to values already reported for Nb contacted GFETs [32]. We also notice that the characteristic measured before the electrical stress shows an asymmetric shape with the *p*-branch clearly away from the expected theoretical behavior. This can be explained in terms of reduced coupling between the Nb electrode and the graphene channel (corresponding to large contact resistance), a situation that can cause asymmetry and/or a double dip in such curves as reported in Ref. [32,33]. The improvement of the contact after electrical stress, resulting in better coupling between Nb and graphene, removed the asymmetry. Comparing the channel resistances, that are extracted as $R_{DS} - R_{contact}$ we confirmed also the improvement of the channel resistance.

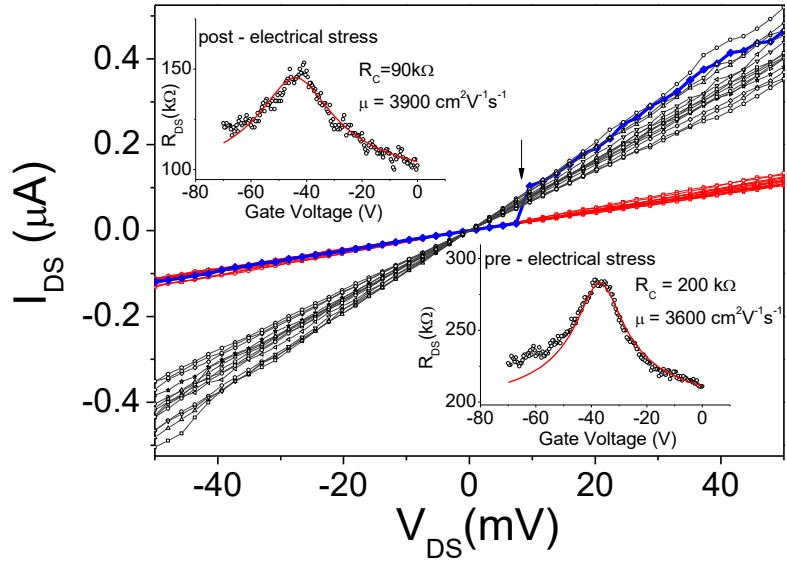

*Figure 2. Output Characteristics ($I_{DS}$ vs $V_{DS}$) and transfer characteristics ($R_{DS}$ vs $V_{Gate}$ in the insets) measured before and after the stabilization of the device due to electrical stress. Black arrow indicates the switch from higher to lower total resistance. Continuous (red) lines in the insets represent the numerical simulations obtained from the model of Ref. [31]. The contact resistance $R_{contact}$ is abbreviated as $R_c$ in the figures.*

In Figure 3 we report the electrical characterization of two devices after stabilization by electrical stress. The curves of figure 3a and 3b are the output characteristics measured in high vacuum (10$^{-7}$ mbar) for different gate voltage values. The ohmic nature of the contacts is confirmed by the linearity of such characteristics.
In figure 3c and 3d we show the corresponding transfer characteristics measured at fixed drain-source bias $V_{DS} = 1$ mV. Remarkably, the current annealing and the long high vacuum storage produced very stable devices with low contact resistance ($5.0\ k\Omega < R_{contact} < 5.5\ k\Omega$). The high fabrication quality is confirmed by the small contact resistance, the low noise and the high carrier mobility that is 4000 V$^2$cm$^{-1}$s$^{-1}$ < μ < 4400 V$^2$cm$^{-1}$s$^{-1}$. The Dirac point at bias between -40 V and -60 V indicates a strong *n*-doping that is favored by the vacuum and the electron irradiation (this measurements was performed inside a SEM, post imaging).

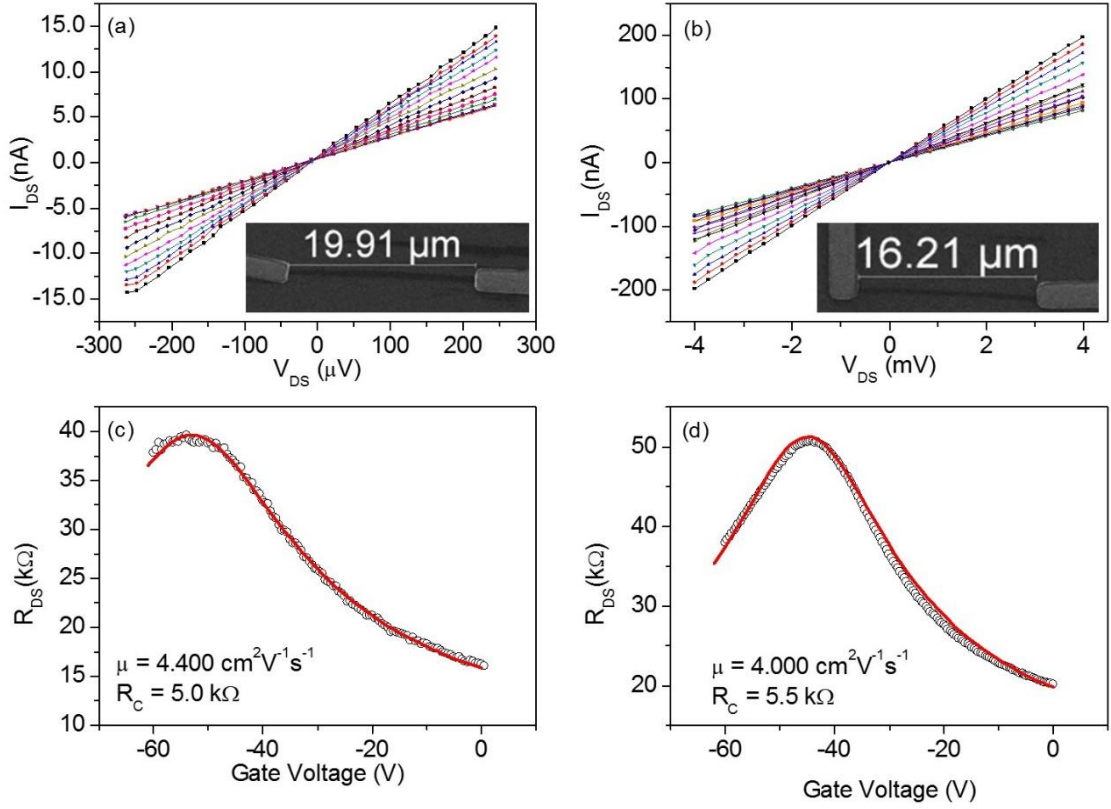

*Figure 3. Electrical characterization under high vacuum of two devices produced on the same substrate. a) and b) $I_{DS}$ vs $V_{DS}$ curves for the devices shown in the insets with dimensions 19,9 μm x 0,7 μm and 16,2 μm x 0,3 μm respectively. c) and d) $R_{DS}$ vs $V_{Gate}$ curves measured at $V_{DS} = 1\ mV$ for the devices of figure 3a and 3b respectively. The solid (red) lines are the fitted model of Ref. [31] with the parameters listed in the plots.*

As soon as the devices are exposed to air, the graphene collects adsorbates that, generally acting as *p*-dopants, shift the Dirac point towards positive biases, increase the contact resistance and reduce the carrier mobility [6, 34-38]. Figure 4a compares the transfer characteristics of the device of figure 3c measured in high vacuum and soon after exposure to air. From the fit of the model, we extracted the contact resistance in air, as $R_{contact} \approx$ 6.7 $k\Omega$, a value 35% larger than the value in high vacuum, while the carrier mobility was reduced to μ ≈ 4100 $V^2cm^{-1}s$. The inset shows the evolution of the Dirac point from -55 V in high vacuum to -30V in air. This observation confirms the importance to perform the electrical measurements *in-situ* when studying irradiation effects, to distinguish electron beam from other environment-induced phenomena.

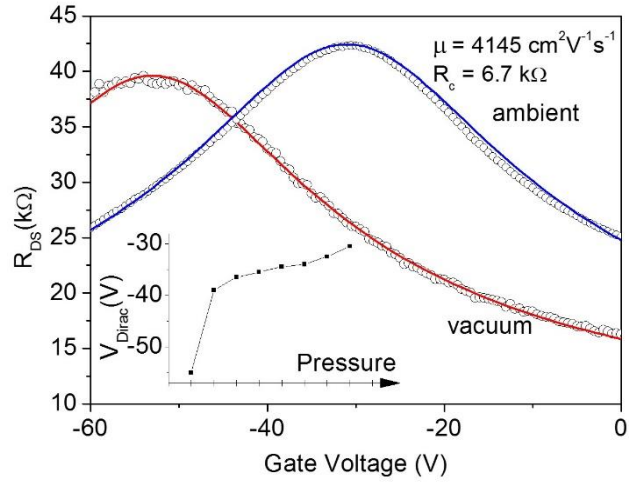

*Figure 4. Effect of the pressure variation from high vacuum to ambient conditions on the $R_{DS}$ vs $V_{Gate}$ curve reported in figure 3a. Solid lines are the fitting curves. In the inset: evolution of the Dirac point for increasing pressure.*

### 3.2 Effect of electron beam irradiation

In the following, we consider the effect of electron beam irradiation on the GFETs. In particular, we consider electron beam energy up to 10 keV, i.e. the energy range typically used for SEM imaging. Larger energy (about 30 keV) is normally used for e-beam lithography or imaging in STEM mode. The irradiation was performed on an area of 20 µm × 20 µm, covering most of the graphene channel, with constant beam current $I_{beam} = 0.2$ nA. We used an exposure time of 10 seconds that resulted in an electron irradiation dose of about 30 e⁻/nm². Differently from other works [26], we performed post-irradiation electrical measurements directly in the SEM chamber, thus avoiding the aforementioned effects of air. Results obtained in six successive electrical sweeps, after a 10 s electron irradiation at 10 keV, are reported in figure 5a. The complete (forward and backward) sweeping between 0 V and -70 V evidences an important hysteresis that decreases with successive electrical sweeps. The appearance of the hysteresis is easily explained by mobile electrons trapped in the gate oxide during e-beam exposure, which screen the gate voltage, while the hysteresis reduction can be caused by their withdrawal by the channel during the successive voltage sweeps [39,40]. By comparing the transfer characteristic before the electron irradiation to the sixth sweep measured after the 10 s exposure (figure 5b), we observe that the device has almost returned to its initial state apart a marginal shift of the Dirac point. To quantitatively analyze the evolution after e-beam exposure (see figure 5c), we used the model of Ref. [31] to estimate the transport parameters, which are summarized in figure 5d. The carrier mobility is significantly reduced by the 10s e-beam irradiation, from 4000 V²cm⁻¹s⁻¹ to about 3600 V²cm⁻¹s (as obtained from the first sweep measurement). The initial value is restored by the successive sweeps. A consistent behavior is shown by the total resistance which is increased by the irradiation and recovers with increasing number of sweeps. The increase of the total resistance, as a consequence of the e-beam irradiation, has been also observed on

CVD grown graphene [41]. Figure 5d reports the effect of irradiation on the contact resistance that is increased of about 70% by the exposure and is smoothly restored by successive sweeps. Noticeably, irradiation seems to have a negligible effect on the intrinsic carrier concentration $n_0$. Mobility and resistance degradation can be explained as increased long-range coulomb scattering [42] by electrons stored in the gate oxide during e-beam exposure (damaging of graphene seems to have a minor contribution); such electrons are gradually removed by voltage application during successive sweeps and pristine conditions are partially recovered.

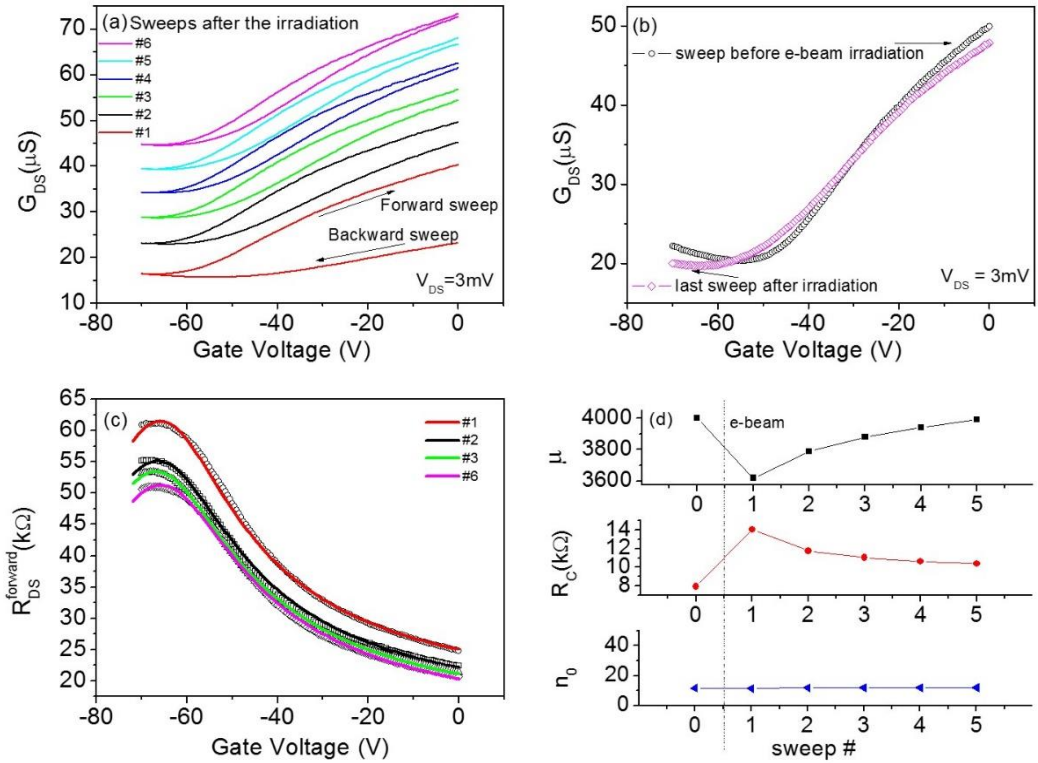

*Figure 5. Effect of electron irradiation on $R_{DS}$ vs $V_{Gate}$ of GFET characterized in figure 3c. a) Six successive sweeps recorded soon after the electron irradiation. Curves have been shifted for clarity; b) Comparison of the sixth sweep after the 10 s e-beam exposure with that measured on unexposed device; c) Forward sweep of selected measurements and relative fitting curves according to the model [31]; d) summary of parameters extracted by fitting of the curves corresponding to forward sweeps.*

## 4. Conclusions.

We realized graphene based field effect transistors on Si/SiO$_2$ substrate with Nb/Au metallic bilayers as contacting electrodes. Electrical characterization evidenced high quality devices with carrier mobility as high as 4000 cm$^2$V$^{-1}$s$^{-1}$ and specific contact resistivity of about 19 k$\Omega\mu$m$^2$. The effect of 10 keV electron irradiation, with dose of 30 e$^-$/nm$^2$, on the transport properties has been reported evidencing a significant reduction of the carrier mobility and an increase of the contact resistance. Finally, we showed that, for low energy irradiation, the pristine conditions are almost restored after several electrical sweeps, which we have explained as gradual removal of electrons piled up in the gate oxide during e-beam exposure.